\documentclass[%
 aip,
 jmp,%
 amsmath,amssymb,
reprint,%
]{revtex4-1}

\usepackage{graphicx}
\usepackage{dcolumn}
\usepackage{bm}
\usepackage{hyphenat}

\begin{document}

\preprint{AIP/123-QED}

\title[Applied Physics Letters xx(2014)]{Energy-efficient magnetoelastic non-volatile memory}

\author{Ayan K. Biswas}

\author{Supriyo Bandyopadhyay}
\affiliation{%
Department of Electrical and Computer Engineering, Virginia Commonwealth University, Richmond, Virginia 23284, USA
}
%

\author{Jayasimha Atulasimha}%
\affiliation{
Department of Mechanical and Nuclear Engineering, Virginia Commonwealth University, Richmond, Virginia 23284, USA
}%

\date{\today}

\begin{abstract}
We propose an improved scheme for low-power writing of binary bits in non-volatile (multiferroic) magnetic memory with electrically generated mechanical stress.
Compared to an earlier idea [Tiercelin, et al., J. Appl. Phys., {\bf 109}, 07D726 (2011)], our scheme improves
distinguishability between the stored bits when
the latter are read with magneto-tunneling junctions. More importantly, the write energy dissipation
and write error rate are reduced significantly
if the writing speed is kept the same.
Such a scheme could be one of the most energy-efficient approaches to writing bits in magnetic non-volatile memory.
\end{abstract}

\maketitle

There is an ongoing quest to devise energy-efficient strategies for writing binary bits in non-volatile magnetic memory.
Writing requires rotating the magnetization of a shape-anisotropic nanomagnet between its two stable orientations that encode the bits `0' and `1'. This
 can be achieved with a magnetic field generated by an electrical current \cite{Alam2010}, a spin transfer torque (STT) arising from a
spin-polarized current \cite{Ralph2008},
or domain wall motion induced by a spin-polarized current \cite{Yamanouchi2004}. A much
more energy-efficient approach is to rotate the magnetization of a two-phase multiferroic
 elliptical nanomagnet, comprising a
magnetostrictive layer in elastic contact with a piezoelectric layer,  with uniaxial mechanical stress
generated by applying an electrical voltage across the piezoelectric layer \cite{Atulasimha2010,Roy2011,Mohammed2011}.
Normally, the maximum rotation possible with such a magneto-elastic scheme
is 90$^{\circ}$,
unless the stress (or voltage) is withdrawn at precisely the right juncture
to allow the magnetization to rotate further to 180$^{\circ}$ \cite{Roy2013}. Such precise withdrawal however is a
challenge, which is why complete bit flips are difficult to achieve. As a result, magneto-elastic switching
 has not been the preferred method to write bits in non-volatile memory, despite its vastly
superior energy-efficiency.

Recently, this impasse was overcome with a clever scheme \cite{Tiercelin2011,Giordano2012,Giordano2013}.
A small in-plane magnetic field is applied along the minor axis of the elliptical magnetostrictive nanomagnet
to move the stable magnetization directions away from the major axis to two mutually perpendicular in-plane directions that lie between the
major and minor axes. They encode the bits `0' and `1'. Uniaxial stress is applied along (or close to) one of these stable directions
(say, the one representing bit `0') by applying an in-plane electric field
between two electrodes delineated
on the pieozelectric layer
(see Fig. 1 of Ref. [\onlinecite{Giordano2012}]). This field generates strain in the piezoelectric layer via the
$d_{33}$ coupling, which is transferred to the magnetostrictive magnet. If the magnet has a positive magnetostriction coefficient, then
tensile stress will rotate the magnetization close to the direction of applied stress
(or electric field) since that orientation will be the global energy minimum. Compressive
stress will rotate it nearly perpendicular to the direction of applied stress, i.e. close to the other stable direction,
since that will become the global
energy minimum. The situation will be the opposite
if the magnetostriction coefficient is negative, but that case is completely equivalent to the first
and hence is not discussed separately. When stress is finally withdrawn, the rotated magnetization will move to the
stable direction closer to the stress-axis,
with $\sim$100\% probability,
and remain there in perpetuity, since that will be energetically favored. Therefore, tensile stress (voltage of one polarity) can be used to write the bit `0' and compressive
stress (voltage of the other polarity) can write the bit `1'. This allows nearly error-free deterministic writing of bits, irrespective
of what the originally stored bit was. A similar idea utilizing 4-state magnets 
was discussed earlier by Pertsev, et al. \cite{pertsev}.

The disadvantage of this scheme is that it restricts the angle between the two stable magnetization orientations to $\sim$$90^\circ$. The stored
bit is usually read with a magneto-tunneling junction (MTJ) that is vertically integrated above or below the magnet. The MTJ will use
the magnetostrictive
magnet
as the soft magnetic layer (or free layer) and a synthetic anti-ferromagnet (SAF) as the hard magnetic layer (or fixed layer) with a tunneling layer
in between. Let us assume that the magnetization of the fixed layer is along the direction that encodes bit `1'. Then the MTJ resistances
with the soft layer's magnetization encoding bit `0' and bit `1' will bear a ratio  $r = {{1 + \eta_1 \eta_2}\over{1 + \eta_1 \eta_2 cos (\Theta)}}$, where the $\eta$-s are the spin injection/detection efficiencies of the two magnet interfaces of the MTJ and $\Theta$ is the
angular separation between the two stable magnetization directions in the MTJ's free layer encoding the two bits.
The maximum value of this ratio (assuming
$\eta_1 = \eta_2 = 1$) is
2:1
since $\Theta \le 90^{\circ}$.
Such a low
ratio may impair the ability to distinguish between bits `0' and `1' in a noisy environment when the bits are read by measuring the MTJ resistance.

We show that the ratio $r$ can be improved without sacrificing any other metric if we introduce {\it two} pairs of electrodes (instead of just
one) to apply
electric fields (and hence stresses)
along two {\it different} directions, each close to a stable magnetization orientation. We will still use a static magnetic field along the
minor axis of the ellipse to displace the stable states from the major axis, but this field will be smaller in strength so
that the displacement from the major axis is smaller.
Consequently,  the angular separation between the stable orientations will be {\it larger} ($\Theta$ $>$ 90$^{\circ}$).
We will need {\it two} pairs of electrodes since merely switching the polarity of the voltage (and hence the sign of the stress)
between any one pair will not switch the magnetization between the two stable states reliably. We shall also apply only {\it one} polarity of
electric field (that always generates compressive stress) between either pair of electrodes. Activating a pair by applying a potential difference
between the corresponding electrodes moves the magnetization by $\sim$90$^{\circ}$ away from the axis joining this pair. Upon deactivation, the magnetization
migrates to the closer stable state with $\geq$ 99.9998\% probability at room temperature and remains there in perpetuity. This writes one bit (say, `0').
If we wish to write the other bit (say, `1'), we will activate the other pair of electrodes.
Similar to the scheme of Refs. [\onlinecite{Tiercelin2011,Giordano2012,Giordano2013}],
this mechanism writes the desired bit with very high reliability ($\geq$ 99.9998\% probability) irrespective of the bit
that was stored earlier in the nanomagnet.

The increased angular separation between the stable orientations immediately increases the
ratio $r$ and improves the distinguishability of the bits. In the rest of this Letter,
we compare our modified scheme with that original scheme of Refs. [\onlinecite{Tiercelin2011,Giordano2012,Giordano2013}] for devices with identical thermal stability factor
\cite{Brown1963}, static error probability and data retention time at room temperature, and switching time. We show that our scheme not only produces a higher ratio $r$, but is
also more energy-efficient and more resilient against dynamic write errors.

\begin{figure}[!ht]
\includegraphics[width=3.4in]{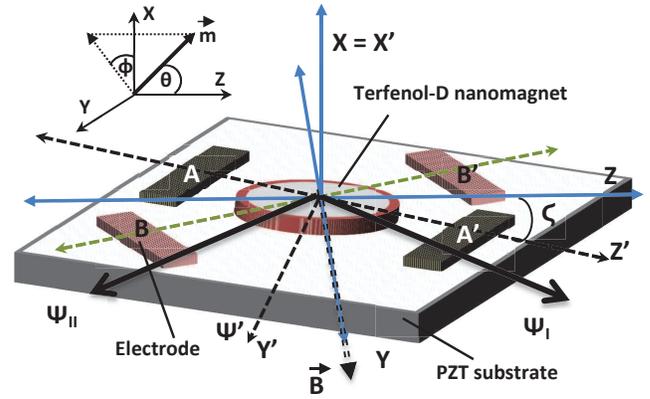}
\caption{\label{fig:fig1} Schematic illustration of the system with two pairs of electrodes (AA$^{\prime}$ and BB$^{\prime}$) and the Terfenol-D
nanomagnet delineated on top of a PZT piezoelectric layer. If the magnetization of the
Terfenol-D nanomagnet was initially in the stable state $\Psi_{I}$ (bit `0'), a voltage applied between the electrode pair
 AA$^{\prime}$ will switch its direction to the other stable state $\Psi_{II}$ (writing the new bit `1'), while a voltage applied
between the pair BB$^{\prime}$ will keep it
 in the original stable state $\Psi_{I}$ (re-writing the old bit `0'). Thus, either bit can be written by activating
the correct electrode pair, irrespective of what the initially stored bit was.}
\end{figure}

Figure \ref{fig:fig1} shows the schematic of our proposed device. The elliptical nanomagnet has a major axis $a$ = 110 nm, minor axis
$b$ = 90 nm, and thickness $d$ = 9 nm. These dimensions ensure that the
nanomagnet has a single magnetic domain \cite{Cowburn1999}. A small magnetic field (B = 8.5 mT) is applied along the in-plane hard axis of the magnet,
 which brings the magnetization stable states out of the major axis, but retain them in the plane of the magnet
($\phi = \pm90^{\circ}$). The new stable states (the two degenerate energy minima) are $\Psi_{I}$ at $\theta =$24.09$^\circ$
 and $\Psi_{II}$ at $\theta =$155.9$^\circ$, where $\theta$ is the angle subtended by the magnetization
vector with the z-axis (or major axis of the elliptical magnet).
 Therefore, the angular separation between these states is $\sim$132$^{\circ}$. The electrodes are delineated such that one pair
 subtends an angle $\zeta=$15$^\circ$ with the z-axis and the other subtends an angle $\zeta=$165$^\circ$. Therefore, the axis joining
 one pair lies close to one stable magnetization direction and the other lies close to the other stable magnetization
direction.

 Application of compressive stress via a voltage applied between the electrode pair AA$^{\prime}$ will write the bit `1', while a
  voltage applied between the electrode pair BB$^{\prime}$ will write `0, irrespective of the initially stored bit.

We define our coordinate system such that the magnet's easy (major) axis lies along the z-axis and the in-plane hard (minor) axis lies along the
y-axis. Uniaxial stress is applied in-plane at an angle $\zeta$ from the easy axis because of the disposition of the electrodes.
To derive general expressions for the
instantaneous potential energies of the nanomagnet due to shape-anisotropy, stress-anisotropy and the static magnetic field, we rotate our coordinate system such that the
z$^{\prime}$-axis in the rotated frame coincides with the direction of applied stress. In the following, quantities with a prime are measured
in the rotated frame of reference.

Using  the rotated coordinate system (see Fig. \ref{fig:fig1}), the shape anisotropy energy of the nanomagnet
$E_{sh}(t)$ can be written as,
\begin{eqnarray}
E_{sh}(t) & = & E_{s1}(t) {\sin}^2{\theta^{\prime}(t)} + E_{s2}(t) {\sin 2\theta^{\prime}(t)} \nonumber\\
          & + & \frac{\mu_{0}}{2} \Omega {M}^2_{s} ( N_{d-yy}{\sin}^2{\zeta} + N_{d-zz} {\cos}^2{\zeta})\nonumber\\
E_{s1}(t) & = & \left( \frac{\mu_{0}}{2} \right) \Omega {M}^2_{s} \{ N_{d-xx}{\cos}^2{\phi^{\prime}(t)} + N_{d-yy}{\sin}^2{\phi^{\prime}(t)} {\cos}^2{\zeta} \nonumber\\
          & - & N_{d-yy} {\sin}^2{\zeta}  + N_{d-zz}{\sin}^2{\phi^{\prime}(t)} {\sin}^2{\zeta}  - N_{d-zz} {\cos}^2{\zeta} \} \nonumber\\
E_{s2}(t) & = & \left( \frac{\mu_{0}}{4} \right) \Omega {M}^2_{s} \left( N_{d-zz} -  N_{d-yy} \right){\sin \phi^{\prime}(t)} {\sin 2\zeta},
\end{eqnarray}
where $\theta^{\prime}(t)$ and $\phi^{\prime}(t)$ are respectively the instantaneous polar and azimuthal angles of the magnetization vector in the
rotated frame,
$M_s$ is the saturation magnetization of the magnet, $N_{d-xx}$, $N_{d-yy}$ and $N_{d-zz}$ are the demagnetization factors
that can be evaluated from the nanomagnet's dimensions \cite{Chikazumi1964}, $\mu_0$ is the permeability of free space,
and $\Omega = (\pi/4)abd$ is the nanomagnet's volume.

The potential energy due to the static magnetic flux density $B$ applied along the in-plane hard axis is given by
\begin{equation}
E_{m}(t) =  M_s \Omega B ({\cos \theta^{\prime}(t)} {\sin \zeta} - {\sin \theta^{\prime}(t)} {\sin \phi^{\prime}(t)} {\cos \zeta} ).
\end{equation}

When a positive voltage is imposed between the electrode pair AA$^{\prime}$, it  generates either compressive or tensile uniaxial stress in the
magnetostrictive nanomagnet depending
on the sign of the magnet's magnetostriction coefficient.  The stress anisotropy energy is given by:
\begin{equation}
E_{str}(t) = - \frac{3}{2}\lambda_{s} \epsilon(t) Y \Omega {\cos}^2{\theta^{\prime}(t)},
\end{equation}
where $\lambda_s$ is the magnetostriction coefficient, $Y$ is the Young's modulus, and $\epsilon(t)$ is the strain generated by the
applied voltage at the instant of time $t$.

The total potential energy of the nanomagnet at any instant $t$ is
\begin{equation}
E(t) = E_{sh}(t) + E_{m}(t) + E_{str}(t).
\end{equation}

\begin{figure}[!ht]
\includegraphics[width=3.5in]{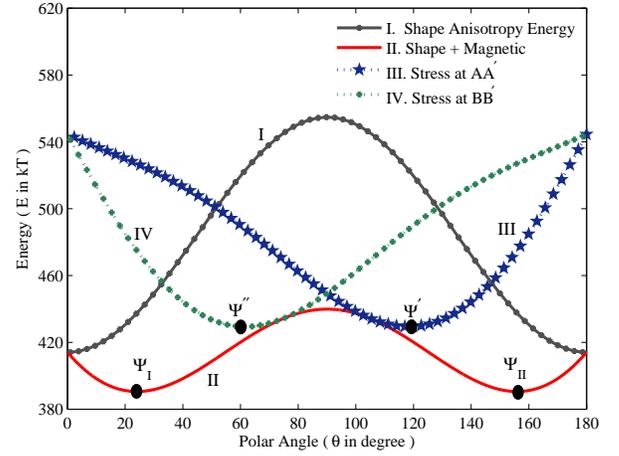}
\caption{\label{fig:fig2} In-plane potential energy profile (azimuthal angle $\phi=$90$^{\circ}$) of the nanomagnet in different conditions.
Curve I shows the profile in the absence of any stress and the static magnetic field, where the energy minima are at
$\theta$ = 0$^{\circ}$, 180$^{\circ}$. Curve II shows the
profile in the presence of an in-plane magnetic field of 8.5 mT along the nanomagnet's minor axis where the energy minima have moved to
 $\theta=$24.09$^{\circ}$ and
at $\theta=$155.9$^{\circ}$. Curve III and IV show the profile when a compressive stress of 9.2 MPa is generated
by imposing a potential between the electrodes AA$^{\prime}$ and the electrodes BB$^{\prime}$ respectively. Note that stress makes the potential profile monostable, instead of bistable.}
\end{figure}

Figure \ref{fig:fig2} shows the potential energy profile of the nanomagnet in the magnet's plane ($\phi$ = 90$^{\circ}$) as a function of the
angle $\theta$ subtended by the magnetization vector with the major axis of the ellipse (z-axis). When no stress is applied
 and the static
magnetic field is absent (curve II), the energy minima and the stable magnetization states lie along the major axis of the ellipse
($\theta$ = 0$^{\circ}$, 180$^{\circ}$) and the in-plane energy barrier separating them is $\sim$145 kT at room temperature.
Application of the static magnetic field along the minor axis (curve I) moves the energy minima and stable magnetization states out of the major axis
to $\theta$ = 24.09$^{\circ}$ and 155.9$^{\circ}$, while reducing the in-plane energy barrier separating the stable states to 49.2 kT. Therefore,
the
probability of spontaneous magnetization flipping between the two stable states due to thermal noise (static error probability)
is $\sim e^{-49.2}$ per attempt \cite{Brown1963},
leading to memory
retention time $(1/f_o)e^{-49.2} = 73$ years, assuming the attempt frequency $f_o$ is 1 THz \cite{Gaunt1977}.
The new stable states are designated as $\Psi_{I}$ (which encodes the binary
bit `0') and $\Psi_{II}$ (which encodes the binary bit `1').

Application of sufficient compressive stress between the electrode pair AA$^{\prime}$ makes the potential profile monostable
(instead of bistable; see curve III) and shifts the  minimum energy position to $\Psi^{\prime}$, so
that the system will go to this state, regardless of whether it was originally at
state $\Psi_I$ or $\Psi_{II}$. After stress removal, the magnetization will end up in the stable state $\Psi_{II}$ (with very high probability at room
temperature) since it is the energy minimum closer to $\Psi^{\prime}$ and getting to $\Psi_{I}$ from
$\Psi^{\prime}$ would have required transcending the
energy barrier between $\Psi^{\prime}$ and $\Psi_{I}$.
Thus, activating the pair AA$^{\prime}$ deterministically writes the bit `1', regardless of the initially stored
bit. Similarly, activating the other pair BB$^{\prime}$ would have written the bit `0' (curve IV of Fig. 2).

In order to calculate the energy dissipated in writing a bit, as well as the probability with which the bit is written
correctly in the presence of thermal noise, we have to solve the stochastic Landau-Lifshitz-Gilbert equation. For this, we proceed
in the standard manner.
The torque that rotates the magnetization in the presence of stress can be written as
\begin{eqnarray}
\mathbf{\tau_{ss}}(t)  & = &  - \mathbf{m}(t) \times \left(\frac{\partial E}{\partial \theta^{\prime}(t)} \hat{\boldsymbol{\theta}} + \frac{1}{\sin \theta^{\prime}(t)}
\frac{\partial E}{\partial \phi^{\prime}(t)} \hat{\boldsymbol{\phi}}\right) \nonumber\\
                      & = & \{E_{\phi 1}(t) \sin\theta^{\prime}(t) + E_{\phi 2}(t) \cos\theta^{\prime}(t) \nonumber\\
                      & - & M_s \Omega B \cos\zeta \cos\phi^{\prime}(t)\}\hat{\boldsymbol{\theta}} \nonumber\\
                      & - & \{E_{s1} (t) \sin2\theta^{\prime}(t) + 2E_{s2} (t) \cos2\theta^{\prime}(t) \nonumber\\
                      & - &  M_s \Omega B (\cos\zeta \sin\phi^{\prime}(t) \cos\theta^{\prime}(t) + \sin\zeta \sin\theta^{\prime}(t))  \nonumber\\
                      & + & (3/2) \lambda_s \epsilon(t) Y \Omega \sin2\theta^{\prime}(t) \}\hat{\boldsymbol{\phi}},
\end{eqnarray}
where $\mathbf{m} (t)$ is the normalized magnetization vector, quantities with carets are unit vectors in the
original frame of reference, and
\begin{eqnarray}
E_{\phi 1}(t) & = & \frac{\mu_{0}}{2} {M}^2_{s} \Omega \{ \left( N_{d-yy} {\cos}^2{\zeta}  + N_{d-zz} {\sin}^2{\zeta} \right) \sin 2\phi^{\prime}(t)  \nonumber \\
              & - & N_{d-xx}\sin 2\phi^{\prime}(t) \} \nonumber \\
E_{\phi 2}(t) & = & \frac{\mu_{0}}{2} {M}^2_{s} \Omega \left( N_{d-zz} - N_{d-yy} \right) \sin 2\zeta \cos \phi^{\prime}(t). \nonumber
\end{eqnarray}

At non-zero temperatures,  thermal noise generates a random magnetic field $\mathbf{h} (t)$ with Cartesian
components $\left ( h_x(t), h_y(t), h_z(t) \right )$
that produces
a random thermal torque which can be expressed as \cite{Roy2012}\\
$\mathbf{\tau_{th}}(t)\! =\! \mu_0 M_s \Omega \mathbf{m}(t)\! \times\! \mathbf{h}(t)\! =\! - \! \mu_0 M_s \Omega \left [ h_{\phi}(t) \hat{\boldsymbol{\theta}} -  h_{\theta}(t) \hat{\boldsymbol{\phi}} \right ]$,
where
\begin{eqnarray}
h_{\theta}(t) & = & h_x (t) \mbox{cos}\theta^{\prime}(t) \cos\phi^{\prime}(t) +  h_y (t) cos\theta^{\prime}(t) \sin\phi^{\prime}(t) \nonumber\\
              & - & h_z (t) sin\theta^{\prime}(t) \nonumber\\
h_{\phi}(t)   & = & -h_x (t) \sin\phi^{\prime}(t) +  h_y (t) cos\phi^{\prime}(t).
\end{eqnarray}

In order to find the temporal evolution of the magnetization vector under the vector sum of the different torques mentioned above,
we solve the stochastic Landau\hyp{}Lifshitz\hyp{}Gilbert (LLG) equation:
\begin{eqnarray}
\frac{d\mathbf{m}(t)}{dt} & - &  \alpha \left[ \mathbf{m}(t) \times  \frac{d\mathbf{m}(t)}{dt} \right]  \nonumber\\
                          & = & \frac{-|\gamma|}{\mu_0 M_s \Omega} \left ( \mathbf{\tau_{ss}}(t)+\mathbf{\tau_{th}}(t) \right )
\end{eqnarray}
From the above equation, we can derive two coupled equations for the temporal evolution of the
polar and azimuthal angles of the magnetization vector:
\begin{eqnarray}
\frac{d\theta^{\prime}(t)}{dt} & = & -\frac{|\gamma|}{(1+\alpha^2)\mu_0 M_s \Omega} \{ E_{\phi 1}(t) \sin\theta^{\prime}(t) + E_{\phi 2}(t) \cos\theta^{\prime}(t) \nonumber\\
                   & - &  M_s \Omega B \cos\zeta \cos\phi^{\prime}(t) - \mu_0 M_s \Omega h_{\phi}(t) \nonumber\\
                   & + & \alpha \{ E_{s1} (t) \sin2\theta^{\prime}(t) - \mu_0 M_s \Omega h_{\theta}(t) \nonumber\\
                   & + & 2E_{s2} (t) \cos2\theta^{\prime}(t) + (3/2) \lambda_s \epsilon(t) Y \Omega \sin2\theta^{\prime}(t) \nonumber\\
                   & - &  M_s \Omega B (\cos\zeta \sin\phi^{\prime}(t) \cos\theta^{\prime}(t) + \sin\theta^{\prime}(t) \sin\zeta ) \} \}
                   \label{eq:finaltheta} \\
\frac{d\phi^{\prime}(t)}{dt}   & = & \frac{|\gamma|}{\sin\theta^{\prime}(t)(1+\alpha^2)\mu_0 M_s \Omega} \{ E_{s1} (t) \sin2\theta^{\prime}(t) \nonumber\\
                   & + & 2E_{s2} (t) \cos2\theta^{\prime}(t) + (3/2) \lambda_s \epsilon(t) Y \Omega \sin2\theta^{\prime}(t) \nonumber\\
                   & - &  M_s \Omega B (\cos\zeta \sin\phi^{\prime}(t) \cos\theta^{\prime}(t) + \sin\zeta \sin\theta^{\prime}(t)) \nonumber\\
                   & - & \mu_0 M_s \Omega h_{\theta}(t) - \alpha ( E_{\phi 1}(t) \sin\theta^{\prime}(t) + E_{\phi 2}(t) \cos\theta^{\prime}(t) \nonumber\\
                   & - &  M_s \Omega B \cos\zeta \cos\phi^{\prime}(t) - \mu_0 M_s \Omega h_{\phi}(t) ) \}. \label{eq:finalphi}
\end{eqnarray}
Solutions of these two equations yield the magnetization orientation $\left ( \theta^{\prime}(t), \phi^{\prime}(t)
\right )$ at any instant of time $t$.

In order to generate the stress-induced magnetodynamics in the presence of thermal noise from the last two equations, we need to pick
(with appropriate statistical weighting) the initial magnetization state from the thermal
distributions around the two stable states $\Psi_I$ and $\Psi_{II}$ in the absence of stress. We determine the
thermal distribution around, say, $\Psi_I$ by
starting with the initial state
$\theta$ = 24.09$^{\circ}$ and $\phi$ =
90$^{\circ}$ and solving Equations (\ref{eq:finaltheta}) and (\ref{eq:finalphi}) to obtain the final values
of $\theta$ and $\phi$
by running the simulation for 1 ns while using a time step of $\Delta t$ = 0.1 ps (the distributions are verified to be independent of
$\Delta t$ and simulation duration). This procedure is then repeated 10$^6$ times to
obtain the thermal distribution of $\theta$ and $\phi$ around $\Psi_I$. The same method is employed to find the
thermal distribution around $\Psi_{II}$.

Let us say that we wish to study the (thermally perturbed) stress-induced magnetodynamics associated with writing the bit `1' when the initial stored
bit was `0'. We apply a voltage between the electrodes $A$ and $A'$ to produce uniaxial stress
and generate a switching trajectory by solving Equations (\ref{eq:finaltheta}) and (\ref{eq:finalphi}) after picking (with appropriate statistical weight) the initial orientation from
the thermal distribution around $\Psi_I$ $( \theta$ = 24.09$^{\circ}$ and $\phi$ = 90$^{\circ} )$ which represents the initial bit `0'.
After the
stress duration is over, the stress is turned off and we continue to simulate the switching trajectory from Equations (\ref{eq:finaltheta}) and (\ref{eq:finalphi}) until the value of
$\theta$ approaches within 4$^{\circ}$ of either $\theta$ = 155.9$^{\circ}$ (correct switching) or $\theta$ = 24.09$^{\circ}$
(failed switching). The switching time is the minimum time needed for nearly all of the
trajectories to switch correctly. It is larger than the stress duration (which is 0.8 ns)
and is about 1.5 ns if 99.9998\%
of the trajectories were to switch correctly.
One million switching trajectories are generated and the fraction of them that fail is the dynamic write error probability.
If no failure occurs, we conclude that the dynamic error probability is less than 10$^{-6}$.

We assume the following material parameters for the magnet (Terfenol-D):
saturation magnetization $M_s = 8\times 10^5$ A/m, magnetostriction coefficient $(3/2)\lambda_s = 90 \times 10^{-5}$, Young's modulus $Y$ = 80 GPa,
and Gilbert damping coefficient $\alpha = 0.1$ \cite{Abbundi1977,Ried1998,Kellogg2008}.
We also assume: strain $\epsilon(t) = 1.15 \times 10^{-4} $ (stress = 9.2 MPa) and $\zeta = 15^{\circ}$.

The $d_{33}$ coefficient of
a bulk PZT substrate
is 3.6$\times$10$^{-10}$ m/V and we assume the same value in a thin film.
 Consequently, in order to generate a strain of
 1.15$\times$10$^{-4}$ in the magnet, one requires an electric field of at least 320 kV/m in the PZT. The voltage
that must be imposed between the electrodes is then 64 mV, assuming the electrode separation to be 200 nm.

The energy dissipated in writing the bit has two components: (1) the {\it internal} dissipation in the nanomagnet due to Gilbert damping,
which is calculated in the manner of Ref. [\onlinecite{Roy2012}] for each trajectory (the mean dissipation
is the dissipation averaged over all trajectories that result in correct switching); and (2) the {\it external} (1/2)$CV^2$ dissipation
associated with applying the voltage across the electrodes which act as a capacitor. Assuming an electrode separation of 200 nm,
substrate thickness of 100 nm, and electrode width of 100 nm, the capacitance is $C$ = 0.44 fF. Therefore, the external
(1/2)$CV^2$ dissipation is 215 kT at room temperature ($V$ = 64 mV). The mean internal dissipation
 could depend on whether the initial stored bit was `0' or `1', and we will take the higher value.
In this case, the higher value was 137 kT.

We found that when the initial stored bit is `0', the bit `1' is written with less than 10$^{-6}$ error probability (not a single
failure among the one million trajectories simulated), while when the initial stored bit is `1', the bit `1' is written
with an error probability of 2$\times$10$^{-6}$ (only two failures among one million trajectories simulated).

Finally, we compare our scheme with that of Ref. [\onlinecite{Tiercelin2011,Giordano2012,Giordano2013}] where compressive or tensile stress is applied at an angle
$\zeta = 45^{\circ}$ with the
major axis of the elliptical nanomagnet to write a bit. In this case, the two stable in-plane magnetization
directions must correspond to $\theta$ = $\sim$45$^{\circ}$ and $\sim$135$^{\circ}$ \cite{Giordano2012} since they must be close to the
stress direction. This would require a higher in-plane static magnetic field since the stable states are to be displaced by a
larger angle from the major axis. We would also want the in-plane barrier height separating the two stable states to be the same
49.2 kT at room temperature. We found that these requirements are satisfied if we choose an elliptical nanomagnet of dimensions
150 nm $\times$ 63 nm $\times$ 11 nm and
 a
static magnetic field (B = 57.3 mT) along the in-plane hard axis. In this case, the stable states are at $\theta = 46^{\circ}$ ($\Psi_{I}$)
and $\theta = 134.5^{\circ}$ ($\Psi_{II}$). The angular separation between the two stable directions is 88.5$^{\circ}$. In order to
get the lowest dynamic error probability in writing a bit, we need to
generate a slightly larger strain of 2.4$\times10^{-4}$ (stress = 19.5 MPa) by applying a slightly larger voltage (135.4 mV).
We also need to keep the strain on for a slightly longer duration (1.5 ns) to complete writing the bit with least 
dynamic error
probability.
With these parameters, we found that the dynamic error probability in writing the bit `1' is 2.1$\times$10$^{-5}$ when the initial
bit is `1' (21 failures in 1 million trajectories) and 5$\times$10$^{-6}$ when the initial
bit is `0' (5 failures in 1 million trajectories). The switching time is still about 1.5 ns. The average internal dissipation is
908 kT (larger because of the larger stress and longer stress duration needed to achieve the same dynamic error probability) and the
external dissipation is 970 kT (larger because of the larger voltage needed to generate the larger stress). The magnet and other
parameters used in Ref. [\onlinecite{Tiercelin2011,Giordano2012,Giordano2013}] were different, but resulted in a much
higher energy dissipation of $\sim$23,000 kT \cite{Giordano2013}. We have therefore re-designed their magnet to reduce the energy dissipation
significantly.

Table \ref{Tab:tab1} presents a comparison between the two schemes where we have assumed that the spin injection and
detection efficiencies ($\eta_1, \eta_2$)
are
$\sim$70\% at room temperature \cite{Salis2005}.

\begin{table}
\caption{Comparison between the 2-electrode and 4-electrode schemes}
\label{Tab:tab1}  
\begin{tabular}{|p{1.9in}|c|c|}
\hline
 & 2-electrode & 4-electrode \\
 \hline
 \centering{Angular separation between stable states ($\Theta$)} & 88.5$^{\circ}$ & 132$^{\circ}$ \\
 \centering{Static error probability at room temperature} & 4.29$\times$10$^{-22}$ & 4.29$\times$10$^{-22}$ \\
 \centering{Dynamic error probability at room temperature} & 2.1$\times$10$^{-5}$ & 2$\times$10$^{-6}$  \\
 \centering{Mean switching time} & 1.5 ns & 1.5 ns \\
 \centering{Mean internal energy dissipation} & 908 kT & 137 kT \\
 \centering{External energy dissipation} & 970 kT & 215 kT \\
 \centering{Mean total energy dissipation} & 1878 kT & 352 kT \\
 \centering{Resistance ratio $r$} & 1.47 & 2.21 \\
 \hline
 \end{tabular}
 \end{table}

In conclusion, we have shown that modifying the scheme of Ref. [\onlinecite{Tiercelin2011,Giordano2012,Giordano2013}] to replace the
single pair of electrodes with two pairs imposes a slight additional lithographic burden, but the payoff in terms of
energy dissipation, dynamic error rate and resistance ratio more than justifies it.
Since the total energy needed to write a bit in the modified scheme is $\sim$350 kT, it could be one of the
most energy-efficient strategies to write bits in non-volatile magnetic memory. Any degradation in the $d_{33}$
coefficient of PZT in a 100-nm thin film will of course require a higher writing voltage and hence a higher amount of
energy dissipation, but since the dissipation is so low, some degradation will be tolerable.

This work was supported by the US National Science Foundation under grants ECCS-1124714 and CCF-1216614. J. A.
would also like to acknowledge the NSF CAREER grant CCF-1253370.


\nocite{*}

\providecommand{\noopsort}[1]{}\providecommand{\singleletter}[1]{#1}%

\end{document}